\documentclass[a4paper,11pt]{article}
\usepackage{jcappub} 
\usepackage{lineno}
\usepackage{bm}
\usepackage{amsfonts}
\usepackage{latexsym}
\usepackage[latin1]{inputenc}
\usepackage{graphicx}
\usepackage{amsmath}
\usepackage{palatino}
\usepackage{textcomp}
\usepackage{float}
\usepackage{booktabs}
\usepackage{dcolumn}
\usepackage{ragged2e}
\usepackage{hyperref}
\hypersetup{colorlinks,citecolor=blue}
\hypersetup{colorlinks=true,linkcolor=blue,filecolor=blue,    urlcolor=magenta}
\usepackage{amsmath}
\usepackage{xcolor}
\usepackage{braket}
\usepackage[caption=false]{subfig}
\usepackage{commath}

\arxivnumber{2302.01564} 
\title{Effects of gravitational lensing by Kaluza-Klein black holes on neutrino oscillations}

\author[a,b]{Hrishikesh Chakrabarty,}
\author[c]{Auttakit Chatrabhuti,}
\author[b]{Daniele Malafarina,}
\author[c]{Bhuddhanubhap Silasan}
\author[d]{and Takol Tangphati} 

\affiliation[a]{School of Astronomy and Space Sciences, University of Chinese Academy of Sciences, Beijing, China}
\affiliation[b]{Department of Physics, Nazarbayev University, 53 Kabanbay Batyr avenue, 010000 Astana, Kazakhstan}
\affiliation[c]{Department of Physics, Faculty of Science, Chulalongkorn University, Bangkok, Thailand}
\affiliation[d]{
School of Science, Walailak University, hasala, Nakhon Si Thammarat, 80160, Thailand
}

\emailAdd{auttakit.c@chula.ac.th}

\abstract{We study gravitational lensing of neutrinos in a Kaluza-Klein black hole spacetime and compare the oscillation probabilities of neutrinos with the case of lensing by black holes in General Relativity. We show that measuring neutrino oscillations in curved spacetimes may allow us to distinguish the two kinds of black holes even in the weak-field limit, as opposed to what happens for the weak lensing of photons. This promises to become an useful tool for future measurements of the properties of black hole candidates and possibly help to constrain the validity of alternative theories of gravity.}


\begin{document}
\maketitle
\flushbottom

\section{Introduction}

With the recent discovery of gravitational waves from binary black hole mergers \cite{LIGOScientific:2016aoc} and observation of the shadow of supermassive black hole candidates \cite{EventHorizonTelescope:2019dse,EventHorizonTelescope:2019ggy,EventHorizonTelescope:2020qrl,EventHorizonTelescope:2022wkp} we are entering an era in which proposed theories of gravity alternative to General Relativity (GR) may be tested via observations \cite{LIGOScientific:2016lio,Will:2014kxa,Yunes:2013dva}. The last few years have seen an enormous increase in the amount of theoretical works aimed at showing how to test the nature of black hole candidates and possible deviations from GR and the observations that are currently providing the most data are:
\begin{itemize}
    \item[(i)] Gravitational waves. So far, LIGO and Virgo have detected about one hundred gravitational waves from merger events of stellar mass black holes \cite{GWTC1,GWTC2,GWTC3,GWTC3-test}. 
    \item[(ii)] Shadow. With  present technology the observation of the shadow is possible only for two supermassive black hole candidates, the one at the center of the galaxy M87 \cite{EventHorizonTelescope:2019dse,EventHorizonTelescope:2020qrl} and the one at the center of the Milky Way \cite{EventHorizonTelescope:2022wkp}.
    \item[(iii)] Accretion disk spectra from Active galactic Nuclei (AGNs). There are thousands of known spectra from quasars and AGNs, which bear information about the properties of the central objects.
    \item[(iv)] Accretion disk spectra from x-ray binaries. Stellar mass black hole candidates in the Milky Way have been detected also from the observation of their accretion disk spectra when they are in a binary system. Again the spectra provide information on the properties of the central object \cite{Bambi:2016sac,Cao:2017kdq,Bambi:2020jpe}.
    \item[(v)] Particle motion in the 'vicinity' of black hole candidates. The motion of stars orbiting a supermassive black hole may be described within the test particle approximation and such orbits have been observed only for the compact object at the Milky Way's center \cite{Ghez:1998ph,Ghez:2003qj,Ghez:2008ms,GRAVITY:2020gka}. 
\end{itemize}
All of these observations may be used to determine some of the properties of black hole candidates. Under the assumption that the central object is a black hole as described in GR it may then be possible to use the observations to measure its mass and spin. On the other hand if the object is described by a black hole mimicker or a black hole in a theory different from GR the same observations may infer different values for the mass, spin and other relevant parameters of the object. 
Many possibilities have been considered in the literature, from wormholes (see for example \cite{Li:2014coa,Nedkova:2013msa}) to modified black hole solutions (see \cite{Abdujabbarov:2016hnw,Cunha:2016wzk,Tsukamoto:2017fxq}), to boson stars and other black hole mimickers (see \cite{Cardoso:2019rvt,Vincent:2015xta,Abdikamalov:2019ztb,Li:2022eue}).

The general picture that is emerging is that one may not be able to rule out alternative theories of gravity and/or black hole mimickers from a single set of observations. Independent measurements of the same property, coming from independent observations are necessary. For this reason it is important to look at other potential observational tools to constrain the geometry in the vicinity of black hole candidates
\cite{Berti:2015itd, Cardoso:2019rvt}.

There have been several theoretical studies on the effects of curved spacetime on neutrino oscillations \cite{Wudka:1991tg,Grossman:1996eh,Cardall:1996cd,Piriz:1996mu,Fornengo:1996ef,Ahluwalia:1996ev,Bhattacharya:1999na,Pereira:2000kq,Crocker:2003cw,Lambiase:2005gt,Godunov:2009ce,Ren:2010yf,Geralico:2012zt,Chakraborty:2013ywa,Visinelli:2014xsa,Chakraborty:2015vla,Zhang:2016deq, Alexandre:2018crg, Blasone:2019jtj, Buoninfante:2019der, Boshkayev:2020igc, Mandal:2021dxk, Koutsoumbas:2019fkn, Swami:2020qdi, Capolupo:2020wlx}. A common conclusion reached in most of these works is the increased effective neutrino oscillation length due to gravitational redshift in curved spacetime backgrounds.  
Inspired from this, one more experimental test that has recently been proposed to potentially break the aforementioned degeneracies from future measurements is the observation of gravitational lensing of neutrinos by massive sources and the effect that it has on neutrino oscillations \cite{Chakrabarty:2021bpr}. Of course, this kind of test is still out of reach from our present technological capabilities. Nevertheless it is important to explore the potential advantages that it may bear over the more 'traditional' lensing of photons. 

It's a very intriguing proposal to test modified theories of gravity using neutrino experiments. We are particularly interested in theoretical models with extra dimensions that were introduced to solve the hierarchy problem between the electroweak and Planck scales as well as inspired by string theory \cite{Arkani-Hamed:1998jmv, Antoniadis:1998ig, Arkani-Hamed:1998sfv}. Indeed, the impact of extra dimensions on neutrino oscillation has been thoroughly investigated in the high energy physics community for over the past two decades. However, most of these investigations are based on the Large Extra Dimensions (LED) model and also make the assumption that sterile neutrinos are present and propagating in a higher-dimensional bulk \cite{Dienes:1998sb, Arkani-Hamed:1998wuz, Barbieri:2000mg} (see \cite{Basto-Gonzalez:2021aus} for recent studies). Here, we provide a different approach that is model independent and does not require the existence of sterile neutrinos to investigate the fingerprints of extra dimensions on the 4-dimensional curved spacetime.


In the present article we consider neutrino oscillations in the geometry of a Kaluza-Klein (KK) black hole \cite{Gibbons:1985ac,Rasheed:1995zv,Larsen:1999pp,Kudoh:2003ki,Horowitz:2011cq}. 
Kaluza-Klein theory is a 5-dimensional extension of GR which includes Maxwell's electrodynamics by considering an additional field in the extra spatial dimension. By compactifying the extra dimension one obtains a 4-dimensional action for which the field equations reduce to those of Einstein and Maxwell. 
Interestingly the KK black hole can also be obtained from a scalar-tensor theory of gravity coupled to Maxwell's electrodynamics with a dilaton field \cite{Frolov:1987rj,Larsen:1999pp}.
The shadow of a KK black hole was studied in \cite{Amarilla:2013sj,Mirzaev:2022xpz} while binary mergers and estimates for the spin and quasinormal modes were considered in \cite{Hirschmann:2017psw,Jai-akson:2017ldo}. More recently, the stability of orbits in the KK black hole was studied in \cite{Blaga:2023exi} while constraints on the theory's parameters from x-ray spectroscopy were obtained in \cite{Tripathi:2021rwb}.
Here we show that the probability of oscillation of neutrinos lensed by a KK black hole and propagating in vacuum towards an observer depends on the geometry via the value of the boost parameter $v$ and thus in principle may allow to distinguish such black holes from the ones in GR. 

The article is organised as follows: A brief review of the KK black hole and neutrino oscillations are provided in sections \ref{KK} and \ref{neutrino}, respectively. In section \ref{4} we develop the formalism to describe neutrino oscillations in the KK metric and in section \ref{5} we obtain the oscillation probabilities for neutrinos lensed by stellar mass KK black holes. 
The implications for future observations of astrophysical black hole candidates are discussed in section \ref{7}.
Throughout the article we make use natural units setting $G=c=1$, with the exception of the plots of the oscillation probabilities where we use GeV units in order to provide more realistic data as expected from lensing by a stellar mass black hole.

\section{The Kaluza-Klein metric}\label{KK}

We consider here the case of the Einstein-Maxwell-dilaton theory arising naturally as a low energy limit in string theory. The action of this theory is essentially that of gravity coupled to Maxwell and a dilaton field \cite{Horne:1992zy}
\begin{equation}
    S=\int d^4x \sqrt{-g}\left[-R+2(\nabla \varphi)^2 + e^{-2\alpha\varphi}F^2\right] ,
\end{equation}
where $R$ is the Ricci scalar that makes for the GR field equations while $F=F_{\mu\nu}F^{\mu\nu}$ is the Faraday tensor of Maxwell's electrodynamics. It has been shown that for $\alpha=\sqrt{3}$ the black hole solution obtained from the above action can be identified with a 5-dimensional KK black hole \cite{Frolov:1987rj,Larsen:1999pp}
and in the following we will restrict our attention to this case. 
In fact, given a 4-dimensional vacuum solution of Einstein's equation in GR and taking its product with $\mathbb{R}$ one
obtains a 5-dimensional solution which is translation invariant in the extra dimension. Then the new solution is obtained by boosting, with boost parameter $v$, the 5-dimensional solution in the new direction. Therefore, to obtain a rotating black hole solution, one starts with the Kerr solution and applies the above procedure. 
When reduced to 4D, the black hole solution obtained can be viewed as a `charged' solution with a non-trivial dilaton field \cite{Dobiasch:1981vh,Chodos:1980df,Gibbons:1985ac}. The resulting Kaluza-Klein line element in Boyer-Lindquist coordinates is given by \cite{Frolov:1987rj}
\begin{equation}\label{metric}
    \begin{aligned}
        ds^2 &= - \frac{1 - Z}{B} dt^2 - \frac{2 a Z \sin^2 \theta}{B \sqrt{1 - v^2}} dt d\phi + \left[ B (r^2 + a^2) + a^2 \sin^2 \theta \frac{Z}{B} \right] \sin ^2 \theta d\phi^2  \\
        & \quad + B \Sigma \left( \frac{dr^2}{\Delta} + d\theta^2 \right),    
    \end{aligned}
\end{equation}
where
\begin{equation}
    \begin{aligned}
        B &= \left( 1 + \frac{v^2 Z}{1 - v^2} \right)^{1/2}, \\
        Z &= \frac{2mr}{\Sigma}, \\
        \Delta &= r^2 + a^2 - 2mr, \\
        \Sigma &= r^2 + a^2 \cos^2 \theta.    
    \end{aligned}
\end{equation}
Here $ m $ and $ a $ are mass and spin parameters of the original Kerr solution and $ v $ is the boost velocity. The dilaton field is given by
\begin{equation}
    \varphi = -\frac{\sqrt{3}}{2}\log B.
\end{equation}
In terms of $m$, $v$, and $a$, the physical mass $M$, charge $Q$, and angular momentum $J$ as measured by observers at infinity are written as
\begin{eqnarray}\label{met-MJQ}
    M &=& m \left( 1 + \frac{v^2}{2(1 - v^2)} \right), \nonumber \\
    Q &=& \frac{mv}{1 - v^2}, \\
    J &=& \frac{ma}{\sqrt{1 - v^2}} \nonumber,
\end{eqnarray}
where the physical mass and charge are not independent anymore since they are related by the boost $v$ which is the result of the reduction of the boosted BH solutions in 5-dimensional gravity into 4-dimensional KK BHs. The locations of the outer and inner horizon of this KK black hole coincide with that of the Kerr solution.

We can see that when $ v \rightarrow 0 $, the metric \eqref{metric} reduces to that of the Kerr solution of GR. On the other hand, $v$ can be interpreted as a boosting parameter with $v\rightarrow 1$ implying that the black hole is boosted at the speed of light. Then to understand the limit $ v \rightarrow 1 $ let us consider the nonrotating case ($ a = 0 $). We can identify the event horizon and the curvature singularity as 
\begin{equation}
    r_{\rm hor} = \frac{2m}{1-v^2}, \ \ \ \ \ r_{\rm sing} = \frac{2mv^2}{1-v^2},
\end{equation}
respectively. For $v\rightarrow 0$ these tend to the values for Schwarzschild. Instead in the extremal limit $v\rightarrow 1$ we have that $ r_{\rm sing} \rightarrow r_{\rm hor} $ remains finite if we take $m\rightarrow 0$ in a suitable way. This corresponds (in the five dimensional viewpoint) to boosting the Schwarzschild solution cross $\mathbb{R}$ to the speed of light $ v = 1 $, while taking the mass parameter $m$ to zero so that the limit is well defined \cite{Horne:1992zy}.
Notice that if we impose that $m/(1-v^2)\rightarrow 0$ in the limit $v\rightarrow 1$ then we have that $B\rightarrow 1$ and the line element \eqref{metric} reduces to Minkowski. Then $M$ and $Q$ go to zero, and no lensing effects are present.  

Looking at the line element \eqref{metric} we can now see that if $a \neq 0$ we can still retrieve a finite limit for $v\rightarrow 1$ if we impose that $m$ goes to zero so that $m/(1-v^2)$ remains finite. In this case we will have that $B\rightarrow \sqrt{1+2c/r}$ is finite and also that $g_{t\phi}\rightarrow 0$. Therefore it is possible to consider the behavior of the metric in the limit of $v\rightarrow 1$ only if we take a vanishing mass parameter. In the following we will restrict the attention to the range $v\in(0,1)$.

\section{Neutrino Oscillations}\label{neutrino}

\begin{figure*}
    \begin{center}
        \includegraphics[width=14cm]{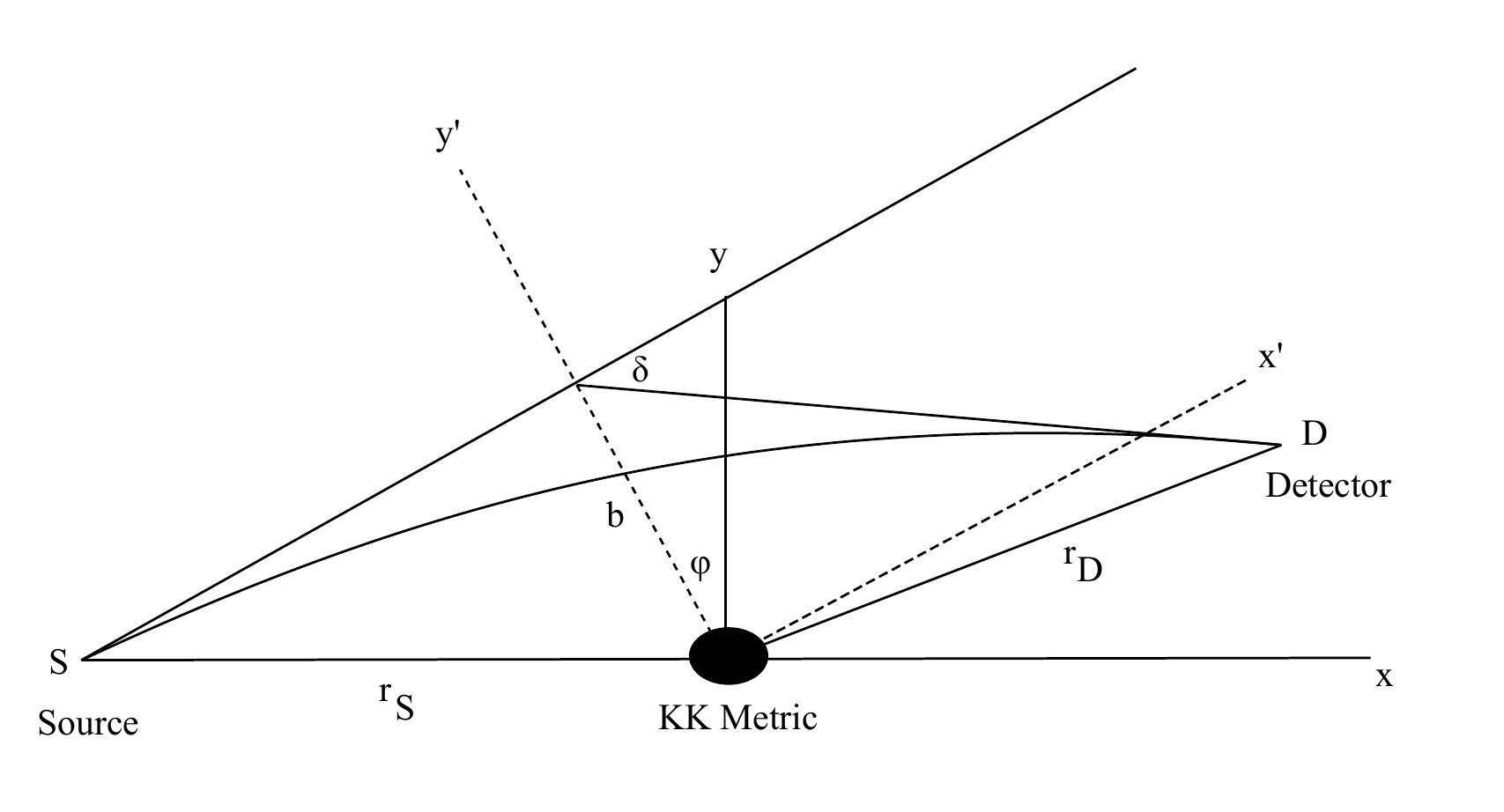}
    \end{center}
    \vspace{-0.5cm}
    \caption{ Schematic diagram for weak lensing of neutrinos in the KK black hole spacetime. Neutrinos propagate from the source $S$ to detector $D$ in the exterior of a static and non-rotating massive object described by the KK-metric. \label{fig1}}
\end{figure*}

Neutrinos are produced and detected in different flavor eigenstates given by $\ket{\nu_\alpha}$, where $\alpha=e,\mu,\tau$. These flavor eigenstates are a superposition of three mass eigenstates denoted by $\ket{\nu_i}$,
\begin{equation}
     \ket{\nu_\alpha} = \sum_{i=1}^3 U^*_{\alpha i}\ket{\nu_i},
\end{equation}
where $i=1,2,3$ and $U^*_{\alpha i}$ is a $3\times3$ unitary mixing matrix \cite{Pontecorvo:1957qd,Pontecorvo:1967fh,Maki:1962mu}. 
Let us assume that the neutrinos have a plane-wave wavefunction and it propagates in vacuum from source $S$ at $(t_S,x_S)$ to detector $D$ at $(t_D,x_D)$ (see Figure \ref{fig1}). So the evolution of the wavefunction at the detector can be given by 
\begin{equation}
    \ket{\nu_i(t_D,x_D)} = \exp(-i\Phi_i)\ket{\nu_i(t_S,x_S)},
\end{equation}
where $\Phi_i$ is the phase of oscillation. Neutrinos produced at the source $ S $ in a flavor eigenstate $ \ket{\nu_\alpha} $ travel to the detector location $ D $ and in this case, the probability of the neutrino changing flavor from the state $\nu_\alpha$ to $\nu_\beta$ is given by
\begin{equation}
    \begin{aligned}
        \mathcal{P}_{\alpha\beta} &= |\braket{\nu_\beta | \nu_\alpha(t_D,x_D)}|^2 = \\
        &= \sum_{i,j}U_{\beta i}U_{\beta j}^*U_{\alpha j}U_{\beta i}^* \exp\left[ -i(\Phi_i - \Phi_j  ) \right].
    \end{aligned}
\end{equation}
In flat spacetime, the phase $ \Phi_i $ has the usual form
\begin{equation}\label{phaseflat}
    \Phi_i = E_i(t_D - t_S) - p_i(\vec{x}_D-\vec{x}_S). 
\end{equation}
In a typical scenario, we assume that the mass eigenstates in a flavour eigenstate initially produced at the source have equal momentum or energy \cite{Bilenky:1978nj}. This assumption, together with $ (t_D-t_S) \simeq |(\vec{x}_D-\vec{x}_S)| $ for relativistic neutrinos ($ E_i \gg m_i $), leads to the phase difference
\begin{equation}
    \Delta \Phi_{ij} \equiv \Phi_i - \Phi_j \simeq \frac{\Delta m_{ij}^2}{2E_0}(\vec{x}_D-\vec{x}_S),
\end{equation}
where $ \Delta m_{ij}^2 = m_i^2 - m_j^2 $, and $ E_0 $ is the average energy of the relativistic neutrinos produced at the source. Therefore, the phenomena of neutrino oscillations in vacuum depend only on the squared mass differences. However, in curved spacetime, the phase and hence the probability depends also on the sum of mass squared of individual mass eigenstates \cite{Swami:2020qdi}. 

In curved spacetime, the expression of the phase $ \Phi_i $ of neutrino propagation can be generalized to a covariant form \cite{Fornengo:1996ef}
\begin{equation}
    \Phi_i = \int^D_S p_\mu^{(i)}dx^\mu,
\end{equation}
where 
\begin{equation}
    p_\mu^{(i)} = m_i g_{\mu\nu}\frac{dx^\nu}{ds}
\end{equation}
is the canonical conjugate momentum to the coordinates $ x^\mu $ and $ g_{\mu\nu} $ and $ ds^2 $ are the metric tensor and line element of the curved spacetime, respectively.

\section{Neutrino Oscillation in KK metric}\label{4}

In this section, we shall calculate the oscillation phases of neutrinos in the KK metric considering the simplest case of a non-rotating object, i.e. setting $ a=0 $. Here we shall follow the notations introduced in \cite{Fornengo:1996ef,Swami:2020qdi,Chakrabarty:2021bpr}. In the equatorial plane $\theta=\pi/2$, the metric components \eqref{metric} become
\begin{equation}\label{gmunu}
    \begin{aligned}
        g_{tt} &= -\left(1-\frac{2m}{r}\right)\left(1+\frac{2mv^2}{r(1-v^2)}\right)^{-1/2}, \\
        g_{rr} &= \left(1-\frac{2m}{r}\right)^{-1}\left(1+\frac{2mv^2}{r(1-v^2)}\right)^{1/2}, \\
        g_{\phi\phi} &= r^2\left(1+\frac{2mv^2}{r(1-v^2)}\right)^{1/2}.
    \end{aligned}
\end{equation}
We also define the components of canonical momenta $p_\mu^{(k)}$ where $k$ corresponds to the mass eigenstate,
\begin{equation}\label{can-mom}
    \begin{aligned}
        p_t^{(k)} &= m_k g_{tt}\frac{dt}{ds}, \\
        p_r^{(k)} &= m_k g_{rr}\frac{dr}{ds}, \\
        p_\phi^{(k)} &= m_k g_{\phi\phi}\frac{d\phi}{ds}.
    \end{aligned}
\end{equation}
The mass shell relation is given by 
\begin{equation}
    -m^2 = g^{tt}p_t^2+g^{rr}p_r^2+g^{\phi\phi}p_\phi^2,
\end{equation}
where we have omitted the index $k$ in order to de-clutter the calculations.

\subsection{Radial case}

Let us first check the neutrino oscillation phase in the radial case, i.e. neutrinos are produced in the gravitational potential and travel outward radially. In this case, $d\phi=0$. Therefore, from Eq.\eqref{can-mom}, we have 
\begin{equation}\label{dtds_drds}
    \frac{dt}{ds}=\frac{p_t}{m_k g_{tt}}, \ \ \ \ \ \ \frac{dr}{ds}=\frac{p_r}{m_k g_{rr}}.
\end{equation}
Since the spacetime is static, we know that the energy $E$ of test particles is conserved, and we define $ p_t = E_k $, $ p_0 = -E_0 $ and $p_r = p_k(r)$, where the subscript $k$ refers to the neutrino flavor. Now, the phase of a neutrino propagating radially in a null trajectory becomes
\begin{equation}\label{phik1}
    \Phi_k = \int_{r_S}^{r_D} \left[-E_k\left(\frac{dt}{dr}\right)_0+p_k(r)\right]dr,
\end{equation}
where the suffix $ 0 $ indicates the null trajectory. Here we would like to emphasize that the phases of neutrino oscillation calculated here would not be the phases on the classical trajectory of the mass eigenstates but the phases calculated on the light-ray trajectory. In the plane wave formalism, this can be taken care of consistently for relativistic neutrinos in the weak-field regime\cite{Stodolsky:1978ks,Bilenky:1978nj}. Readers are also referred to the appendix of Ref.\cite{Swami:2020qdi} for a proof. Now from Eq.\eqref{dtds_drds}
\begin{equation}\label{dtdr0}
    \left(\frac{dt}{dr}\right)_0 = -\frac{E_0}{p_0(r)}\frac{g_{rr}}{g_{tt}},
\end{equation}
on a null trajectory, where $E_0$ is the energy of the massless particle at infinity and we have dropped suffix $k$ from $ p_{k,0}(r) $ to avoid clutter. The mass-shell relation now gives
\begin{equation}\label{p0pr}
    p_0 = \pm E_0\sqrt{-\frac{g_{rr}}{g_{tt}}}, \ \ \ \ \ p_k(r) = \pm \sqrt{-\frac{g_{rr}E_k^2}{g_{tt}}-m_k^2 g_{rr}}.
\end{equation}
Now using Eq.\eqref{dtdr0} and Eq.\eqref{p0pr}, we can write the phase in the null trajectory Eq.\eqref{phik1} as
\begin{equation}
    \Phi_k = \pm\int_{r_S}^{r_D}E_k \sqrt{-\frac{g_{rr}}{g_{tt}}}\left[\sqrt{1-\frac{m_k^2g_{tt}}{E_k^2}}-1\right]dr.
\end{equation}
Since $ 0 < g_{tt} < 1 $, we can expand the square root inside the bracket to get
\begin{equation}\label{phik2}
    \Phi_k = \pm \int_{r_S}^{r_D}\sqrt{-g_{tt}g_{rr}}E_k\frac{m_k^2}{2E_k^2}dr.
\end{equation}
Now, in relativistic approximation ($m_k<<E_k$), the following expressions hold
\begin{equation}
    \begin{aligned}
        &E_k \simeq E_0 +\mathcal{O}\left(\frac{m_k^2}{2E_0}\right), \\
        &E_k\frac{m_k^2}{2E_k^2} \simeq E_0 \frac{m_k^2}{2E_0^2}.
    \end{aligned}
\end{equation}
So using these two relations, Eq.\eqref{phik2} becomes
\begin{equation}
    \Phi_k = \pm \frac{m_k^2}{2E_0} \int_{r_S}^{r_D} \sqrt{-g_{tt}g_{rr}}dr.
\end{equation}
Now for the metric coefficients in Eq.\eqref{gmunu}, we have $ g_{tt}g_{rr} = -1 $. So the phase becomes
\begin{equation}
    \Phi_k = \pm \frac{m_k^2}{2E_0} (r_D - r_S),
\end{equation}
which has the same form as that in a flat spacetime. Hence, the oscillation probability of neutrinos travelling radially outward in the equatorial plane of a weak gravitational field will be independent of the gravitational effects. This independence can be understood as the consequence of spherical symmetry and the restriction of neutrino travel only to the equatorial plane. In fact the same will hold in any spacetime where $ g_{tt}g_{rr} = -1 $. 

\subsection{Non-radial case}

Let us now investigate what happens in the non-radial case. For a neutrino travelling along a null trajectory, the phase can be written using Eq.\eqref{can-mom} and Eq.\eqref{p0pr} as
\begin{equation}
    \Phi_k = \int_{r_S}^{r_D}\left[ -E_k \left( \frac{dt}{dr} \right)_0 + p_r + J_k \left( \frac{d\phi}{dr} \right)_0 \right]dr,
\end{equation}
where we use conservation of energy and angular momentum $ p_t^{(k)}=-E_k $ and $ p_\phi^{(k)} = J_k $ and $ J $ is the angular momentum of the test particle. Now, again from Eq.\eqref{can-mom}, we can write
\begin{equation}
    \frac{dt}{dr} = \frac{-E_k}{p_r}\frac{g_{rr}}{g_{tt}}, \ \ \ \ \ \ \frac{d\phi}{dr} = \frac{J_k}{p_r}\frac{g_{rr}}{g_{\phi\phi}}.
\end{equation}
Along the null trajectory, these equations take the form
\begin{equation}\label{nr-dtdr}
    \left(\frac{dt}{dr}\right)_0 = -\frac{E_0}{p_0(r)}\frac{g_{rr}}{g_{tt}}, \ \ \ \ \ \ \ \left(\frac{d\phi}{dr}\right)_0 = \frac{J_0}{p_0(r)}\frac{g_{rr}}{g_{\phi\phi}}.
\end{equation}
Now, we would like to express the angular momentum $J_k$ as a function of the energy $E_k$, the impact parameter $b$ and the velocity of the test particle at infinity $v_k^{(\infty)}$ as
\begin{equation}\label{nr-jk}
    J_k = E_k b v_k^{(\infty)}.
\end{equation}
The metric is asymptotically flat, so using relativistic approximation, we can express the velocity and the angular momentum as
\begin{equation}\label{nr-jkvk}
    \begin{aligned}
        &v_k^{\infty} = \frac{\sqrt{E_k^2-m_k^2}}{E_k} \simeq 1 - \frac{m_k^2}{2E_k^2}, \\
        &J_k  \simeq E_k b \left( 1 - \frac{m_k^2}{2E_k^2} \right),
    \end{aligned}
\end{equation}
and for a massless particle
\begin{equation}\label{nr-j0}
    J_0 = E_0 b.
\end{equation}
Using Eqs. \eqref{nr-dtdr}, \eqref{nr-jk}, \eqref{nr-jkvk} and \eqref{nr-j0}, we can write the phase for the non-radial case as
\begin{equation}
    \begin{aligned}
        \Phi_k = \int_{r_S}^{r^D}\frac{E_0 E_k g_{rr}}{p_0(r)} &\Bigg[ \frac{1}{g_{tt}} + \frac{p_0 p_k(r)}{E_0 E_k g_{rr}} +\frac{b^2}{g_{\phi\phi}}\left(1 - \frac{m_k^2}{2E_k^2}\right) \Bigg]dr.
    \end{aligned}
\end{equation}
Now, we shall simplify this expression using the mass-shell relation. The following two equations can be obtained from the mass-shell relation
\begin{equation}\label{nr-msrels}
    \begin{aligned}
        \frac{p_0(r)}{E_0 g_{rr}} &= \pm \sqrt{-\frac{1}{g_{tt}g_{rr}} - \frac{b^2}{g_{rr}g_{\phi\phi}}}, \\
        \frac{p_0(r)p_k(r)}{E_0 E_k B} &= -\left( \frac{1}{g_{tt}} + \frac{b^2}{g_{\phi\phi}} + \frac{m_k^2}{2E_k^2} \right),
    \end{aligned}
\end{equation}
And from these, we can express the phase as
\begin{equation}
    \Phi_k = -\int_{r_S}^{r_D} \frac{E_0 g_{rr}}{p_0(r)} E_k \frac{m_k^2}{2E_k^2}dr. 
\end{equation}
In the relativistic approximation, we get
\begin{equation}
    \begin{aligned}
        \Phi_k &= \frac{m_k^2}{2E_0}\int_{r_S}^{r_D}\frac{E_0 g_{rr}}{p_0(r)}dr = \\
        &= \frac{m_k^2}{2E_0}\int_{r_S}^{r_D}\sqrt{-g_{tt}g_{rr}}\left(1-\frac{b^2|g_{tt}|}{g_{\phi\phi}}\right)^{-1/2}dr.
    \end{aligned}
\end{equation}
We can see that for $ b = 0 $, we recover the usual phase expression flat background spacetime. However, for $b\neq 0$ we see that the geometry now plays a role. 

Now we shall evaluate the integral in two cases. 
\begin{itemize}
    \item[(i)] Neutrinos are produced in the gravitational potential and travel outwards non-radially towards the detector.
    \item[(ii)] Neutrinos coming from a distant source are lensed by the gravitational potential before reaching the detector.
\end{itemize} 
Let us consider the integrand in the first case. Replacing the metric coefficients, we get
\begin{equation}
    \begin{aligned}
        \sqrt{-g_{tt}g_{rr}}\left( 1 - \frac{b^2|g_{tt}|}{g_{\phi\phi}} \right)^{-1/2} &= \left[ \frac{b^2}{r^2}\left( \frac{2m}{v^2(2m-r)+r} - 1 \right) + 1 \right]^{-1/2} \\
        &= \left[ \frac{b^2}{r^2}\left( \frac{4M}{v^2(4M-r)+2r} - 1 \right) + 1 \right]^{-1/2}.
    \end{aligned}
\end{equation}
Here, in the second equality, we have used Eq.\eqref{met-MJQ} to express the integrand in terms of the physical mass $M$. We now apply the weak field limit $ M/r << 1 $ to obtain
\begin{equation}
    \begin{aligned}
        \left[ \frac{b^2}{r^2}\left( \frac{4M}{v^2(4M-r)+2r} - 1 \right) + 1 \right]^{-1/2} &= \left( \frac{r^2}{r^2-b^2} \right)^{1/2} + \frac{2 b^2 M}{(v^2 - 2)(r^2 - b^2)^{3/2}} + \mathcal{O}(M^2).
    \end{aligned}
\end{equation}
Now we perform the integral to obtain
\begin{equation}\label{eq-intfull}
    \begin{aligned}
        \int \sqrt{-g_{tt}g_{rr}}\left(1-\frac{b^2|g_{tt}|}{g_{\phi\phi}}\right)^{-1/2}dr &= \sqrt{r^2-b^2} + \frac{Mr}{\sqrt{r^2-b^2}} -\frac{M r v^2}{(v^2 - 2)\sqrt{r^2-b^2}}.
    \end{aligned}
\end{equation}
The first two terms here are the Schwarzschild contribution and the third term arises because of the non-zero boost velocity parameter. This term can also be written in terms of the charge as
\begin{equation}
    -\frac{M r v^2}{(v^2 - 2)\sqrt{r^2-b^2}} = \frac{Q r v}{2\sqrt{r^2 - b^2}}.
\end{equation}
Now by solving the relation $ Q/M = 2v/(2-v^2) $, we can find that the parameter $ v $ has two real solutions
\begin{equation}
    v = \frac{\pm \sqrt{M^2 + 2Q^2} - M}{Q}.
\end{equation}
For small physical charge, $ Q/M << 1 $, the positive part yields $ |v| \sim \left| Q/M \right| << 1 $. On the other hand, the negative part yields $ |v| \sim 2\left| M/Q \right| >> 1 $, which is nonphysical. Going forward, we shall only consider the positive part of the solution. This gives
\begin{equation}
    \begin{aligned}
        \int_{r_S}^{r_D}\sqrt{-g_{tt}g_{rr}}\left(1-\frac{b^2|g_{tt}|}{g_{\phi\phi}}\right)^{-1/2}dr &= \sqrt{r^2-b^2} + \frac{Mr}{\sqrt{r^2-b^2}} + \frac{Q^2 r}{2M\sqrt{r^2 - b^2}} \\
        &= \sqrt{r^2-b^2} + M\left( 1 + \frac{Q^2}{2M^2} \right)\frac{r}{\sqrt{r^2 - b^2}}.
    \end{aligned}
\end{equation}
Then, the phase can be written as
\begin{equation}
    \begin{aligned}
        \Phi_k &= \frac{m_k^2}{2E_0}\Bigg[ \sqrt{r_D^2 - b^2} - \sqrt{r_S^2 - b^2} + M \left( 1 + \frac{Q^2}{2M^2} \right) \left( \frac{r_D}{\sqrt{r_D^2 - b^2}} - \frac{r_S}{\sqrt{r_S^2 - b^2}} \right)  \Bigg].
    \end{aligned}
\end{equation}
In the limit $ v \rightarrow 0 $, the above expression reduces to that of the Schwarzschild case \cite{Fornengo:1996ef,Swami:2020qdi}. 

Now, let us consider the second case. Here, neutrinos are produced from a distant source, travel towards the lensing object, passing at closest approach at $ r = r_0 $ and subsequently reaching the detector. The phase, in this case, can be written in two parts taking into account the sign of the momentum and the symmetry of the trajectory about $r_0$. Then
\begin{equation}
    \begin{aligned}
        \Phi_{k}(r_S \rightarrow r_0 \rightarrow r_D) &= \frac{m_k^2}{2E_0}\int_{r_0}^{r_S}\sqrt{\frac{-g_{tt}g_{rr}}{1-\frac{b^2|g_{tt}|}{g_{\phi\phi}}}}dr + \frac{m_k^2}{2E_0}\int_{r_0}^{r_D}\sqrt{\frac{-g_{tt}g_{rr}}{1-\frac{b^2|g_{tt}|}{g_{\phi\phi}}}}dr.
    \end{aligned}
\end{equation}
In the weak-field approximation, the point of closest approach can be determined by solving the equation
\begin{equation}
    \left( \frac{dr}{d\phi} \right)_0 = \frac{p_0(r_0)g_{\phi\phi}}{J_0 g_{rr}} = 0.
\end{equation}
The solution of this equation takes a simple form and is given by
\begin{equation}
    r_0 \simeq b - \frac{2M}{2-v^2} \simeq b - M\left( 1 + \frac{Q^2}{2M^2} \right).
\end{equation}
In the second equality, we considered a small charge $ v \sim Q/M << 1 $. Then we express the integrand of the phase in terms of $ r_0 $ in the weak field limit as
\begin{equation}
    \begin{aligned}
        \sqrt{\frac{-g_{tt}g_{rr}}{1-b^2|g_{tt}|/g_{\phi\phi}}} = \frac{r}{\sqrt{r^2 - r_0^2}} - \frac{2 M r_0}{(v^2 - 2)(r+r_0)\sqrt{r^2 - r_0^2}} + \mathcal{O}(M^2/r^2).
    \end{aligned}
\end{equation}
Then we perform the integral to get
\begin{equation}
    \begin{aligned}
        \int_{r_S}^{r_D}\sqrt{-g_{tt}g_{rr}}\left(1-\frac{b^2|g_{tt}|}{g_{\phi\phi}}\right)^{-1/2}dr &= \sqrt{r^2-r_0^2} - \frac{2M\sqrt{r^2-r_0^2}}{(r+r_0)(v^2-2)}  \\
        &= \sqrt{r^2-r_0^2} + M\left( 1 + \frac{Q^2}{2M^2} \sqrt{\frac{r-r_0}{r+r_0}} \right),
    \end{aligned}
\end{equation}
where in the second equality, we have considered the small charge approximation $ v \sim Q/M << 1 $. 
Finally, we obtain
\begin{equation}
    \begin{aligned}
        \Phi_k (r_S \rightarrow r_0 \rightarrow r_D) &= \frac{m_k^2}{2E_0}\Bigg[ \sqrt{r_D^2 - r_0^2} +\sqrt{r_S^2 - r_0^2} \\ 
        &\quad + M\left( 1 + \frac{Q^2}{2M^2} \right) \left( \sqrt{\frac{r_D - r_0}{r_D + r_0}} + \sqrt{\frac{r_S - r_0}{r_S + r_0}} \right) \Bigg], \\
    \end{aligned}
\end{equation}
Or
\begin{equation}
    \begin{aligned}
        \Phi_k &= \frac{m_k^2}{2E_0}\Bigg[ \sqrt{r_D^2 - b^2} +\sqrt{r_S^2 - b^2} + M\left( 1 + \frac{Q^2}{2M^2} \right) \Bigg( \frac{b}{\sqrt{r_D^2 - b^2}} + \frac{b}{\sqrt{r_S^2 - b^2}} \\
        &\quad +\sqrt{\frac{r_D-b}{r_D + b}} + \sqrt{\frac{r_S-b}{r_S + b}}\Bigg)\Bigg].
    \end{aligned}
\end{equation}
Now, for $ b << r_{S,D} $, we expand the above expression over $ b/r_{S,D} $ and keep terms up to $ \mathcal{O}\left(b^2/r_{S,D}^2\right) $ to get
\begin{equation}\label{eq-phasej24}
    \begin{aligned}
        \Phi_k = \frac{m_k^2}{2E_0}(r_D + r_S) \Bigg[ 1 - \frac{b^2}{2 r_D r_S} + \left( 1 + \frac{Q^2}{2M^2} \right)\frac{2M}{r_D + r_S} \Bigg].
    \end{aligned}
\end{equation}
So, we can see that the phase depends on the boost parameter through the charge $ Q $. When $ v \rightarrow 0 \ \ (Q \rightarrow 0) $ we recover the expression for the phase in the Schwarzschild spacetime \cite{Fornengo:1996ef}.

If one wishes to consider the case of $v\rightarrow 1$ then equation \eqref{eq-intfull} must be used. Here we only notice that the probability of neutrino oscillations in this case will not contain any divergences as long as we take the limit according to the prescription given in section \ref{KK}, namely $m\rightarrow 0$ as $v\rightarrow 1$ in such a way that $m/(1-v^2)\rightarrow {\rm const.}$. In this case both $Q$ and $M$ are finite for $v\rightarrow 1$ and also $Q/M\rightarrow 2$. Also, if we have $m/(1-v^2)\rightarrow 0$ then $M=0$ and the spacetime reduces to Minkowski. Then there will be no gravitational effects on neutrino oscillation, as it can be seen from equation \eqref{eq-phasej24}.

\begin{figure}
    \begin{center}
        \includegraphics[width=8.5cm]{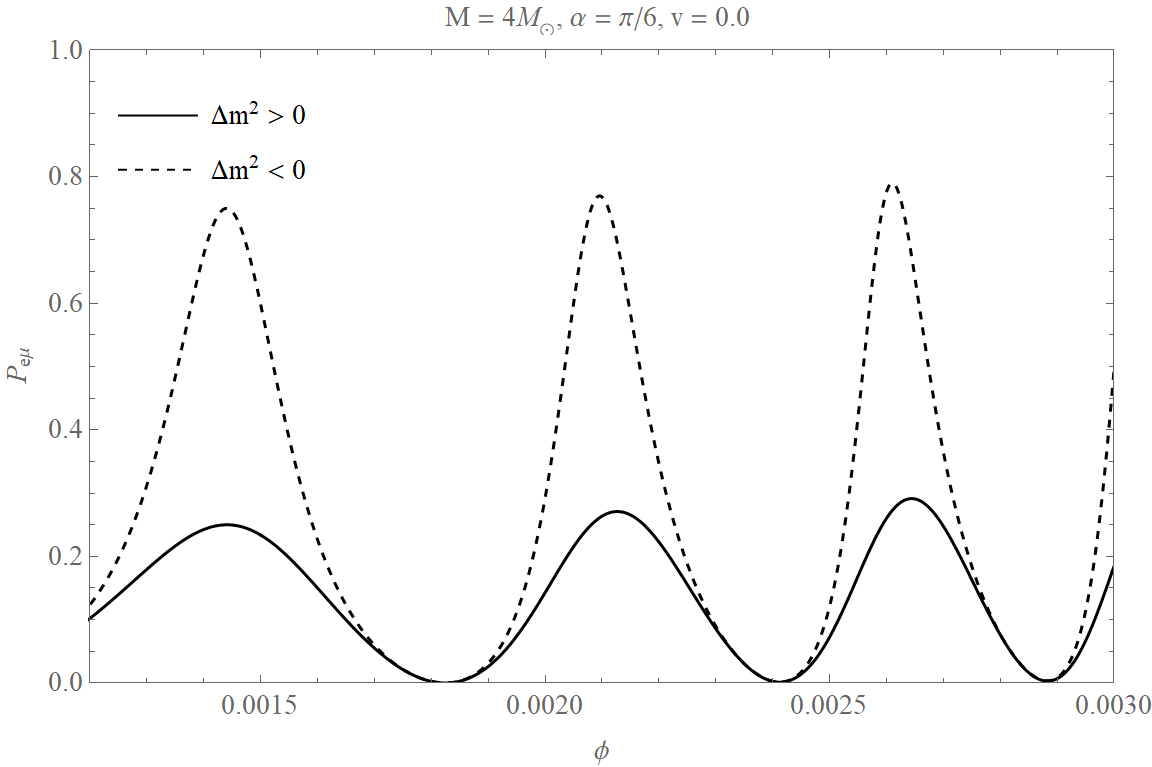}
    \end{center}
    \caption{ Neutrino oscillation probability for $ v = 0 $, i.e. in the Schwarzschild case, when the mixing angle is $ \pi/6 $. The solid and the dashed curves represent normal hierarchy and inverted hierarchy respectively. Values of the other parameters are as follows: $ M = 4M_\odot $, $ \left| \Delta m^2 \right| = 10^{-3} \rm{eV}^2 $, and the lightest neutrino is massless. We use Eq.~\eqref{eq-pr} and \eqref{eq-norm} with \eqref{aijbpq} to plot this figure.  \label{fig2}}
\end{figure}

\section{Neutrino oscillation probability}\label{5}

\begin{figure*}
    \begin{center}
        \includegraphics[width=0.45\textwidth]{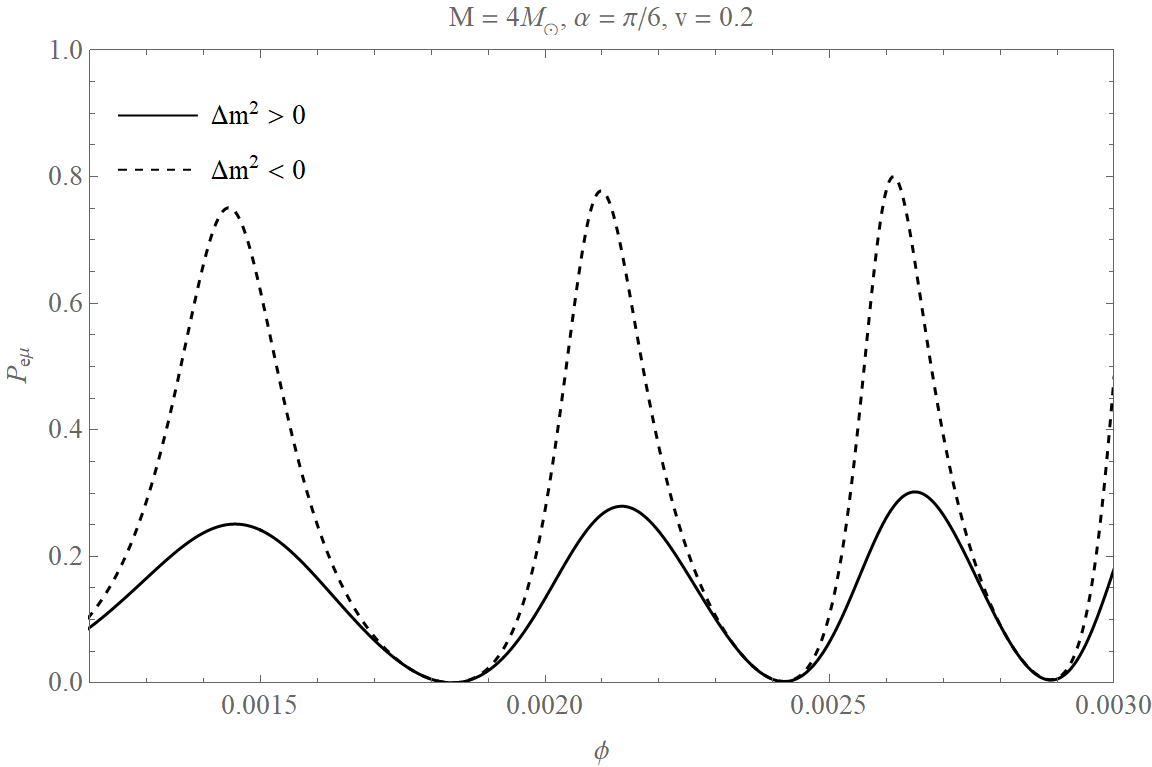}
        \includegraphics[width=0.45\textwidth]{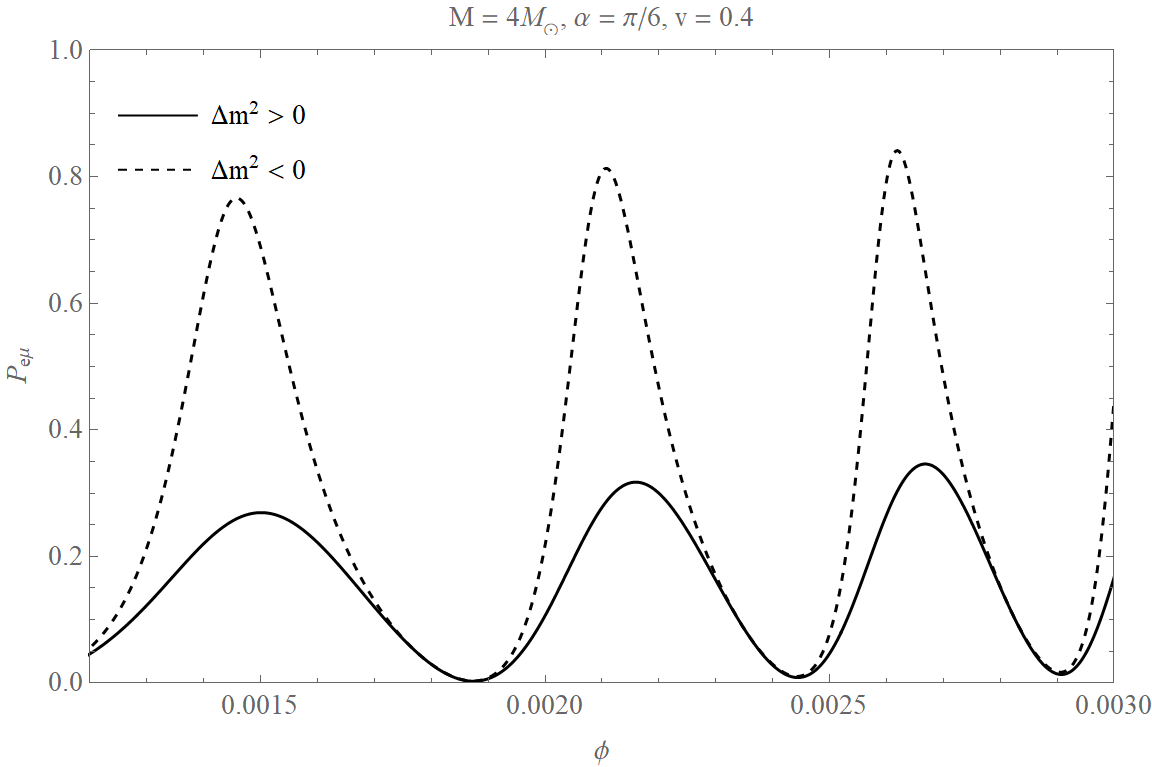}
    \end{center}
    \caption{ Left Panel: Neutrino oscillation probability for $ v = 0.2 $ when the mixing angle is $ \pi/6 $. Right panel: Neutrino oscillation probability for $ v = 0.4 $ when the mixing angle is $ \pi/6 $. The solid and the dashed curves represent normal hierarchy and inverted hierarchy respectively. Values of the other parameters are as follows: $ M = 4M_\odot $, $ \left| \Delta m^2 \right| = 10^{-3} \rm{eV}^2 $, and the lightest neutrino is massless. We use Eq.~\eqref{eq-pr} and \eqref{eq-norm} with \eqref{aijbpq} to plot these figures. 
 \label{fig3}}
\end{figure*}

\begin{figure*}
    \begin{center}
        \includegraphics[width=0.45\textwidth]{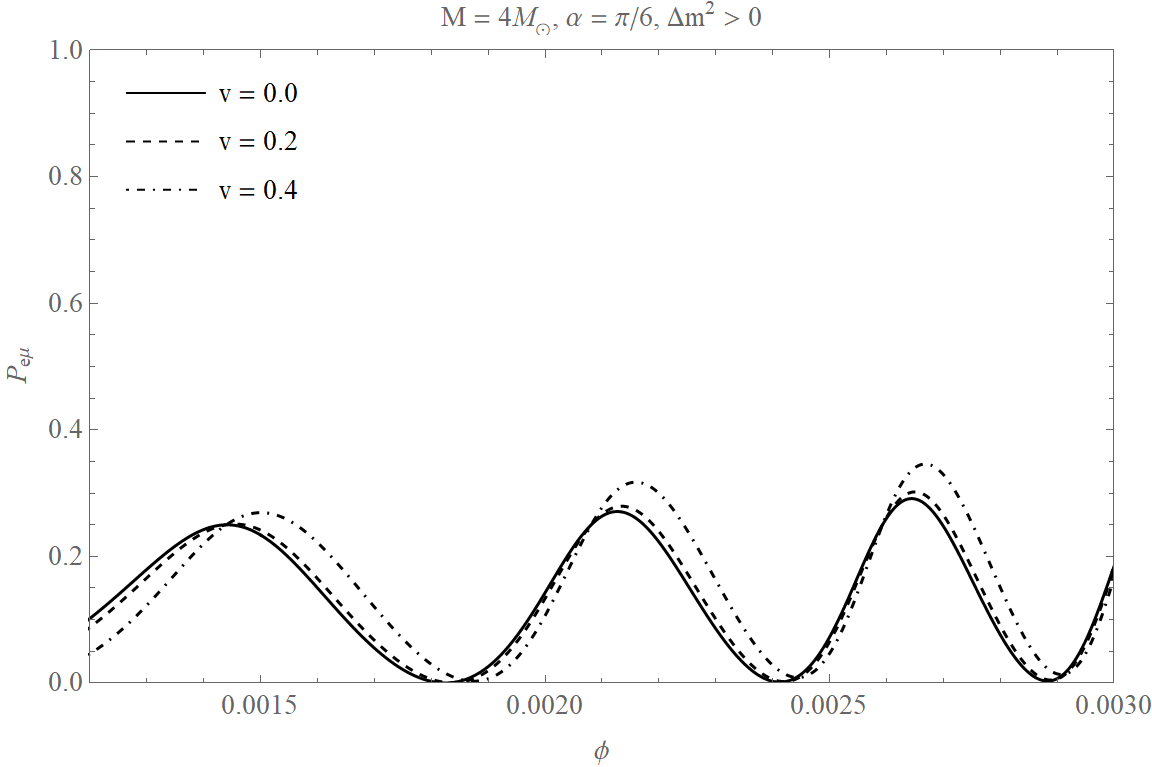}
        \includegraphics[width=0.45\textwidth]{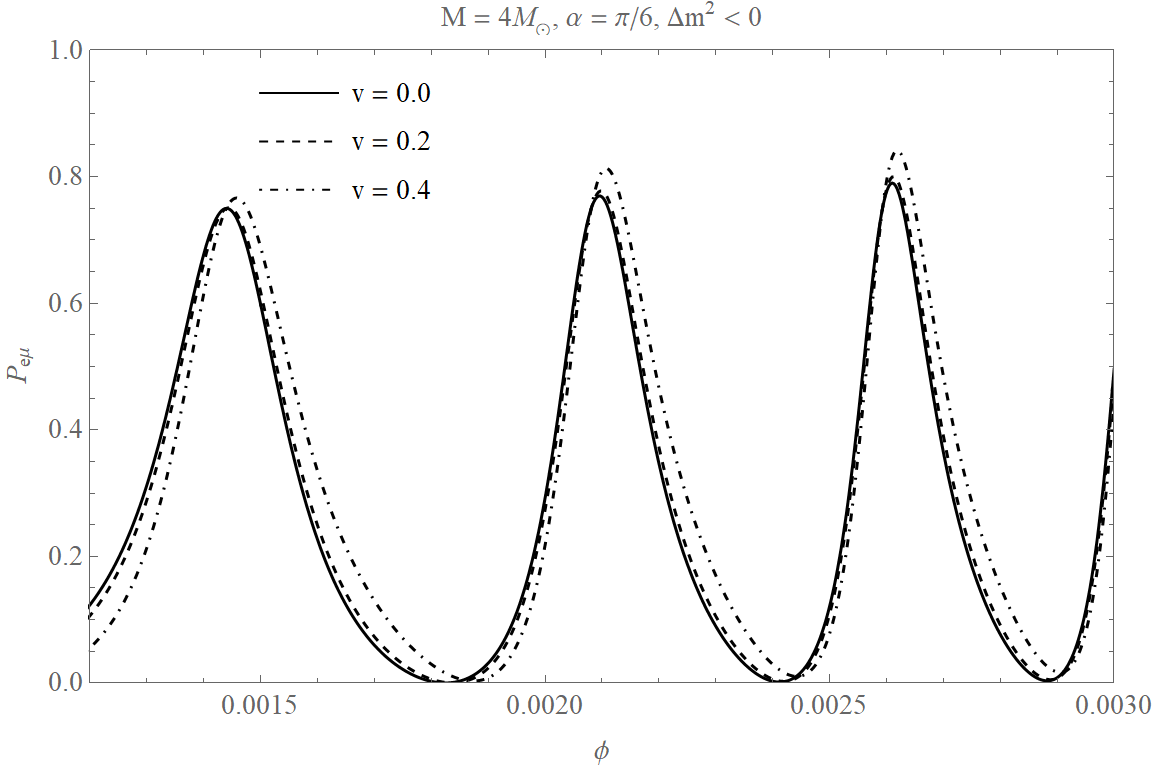}
    \end{center}
    \caption{ Left Panel: Neutrino oscillation probability for $ v = 0 $ (solid curve), $ v = 0.2 $ (dashed curve) and $ v = 0.4 $ (dotted-dashed curve) for normal hierarchy. Right panel: Neutrino oscillation probability for $ v = 0 $ (solid curve), $ v = 0.2 $ (dashed curve) and $ v = 0.4 $ (dotted-dashed curve) for inverted hierarchy. 
    Values of the other parameters are as follows: $ M = 4M_\odot $, $ \left| \Delta m^2 \right| = 10^{-3} \rm{eV}^2 $, $ \alpha = \pi/6 $ and the lightest neutrino is massless. We use Eq.~\eqref{eq-pr} and \eqref{eq-norm} with \eqref{aijbpq} to plot these figures. \label{fig4}}
\end{figure*}

\begin{figure*}
    \begin{center}
        \includegraphics[width=0.45\textwidth]{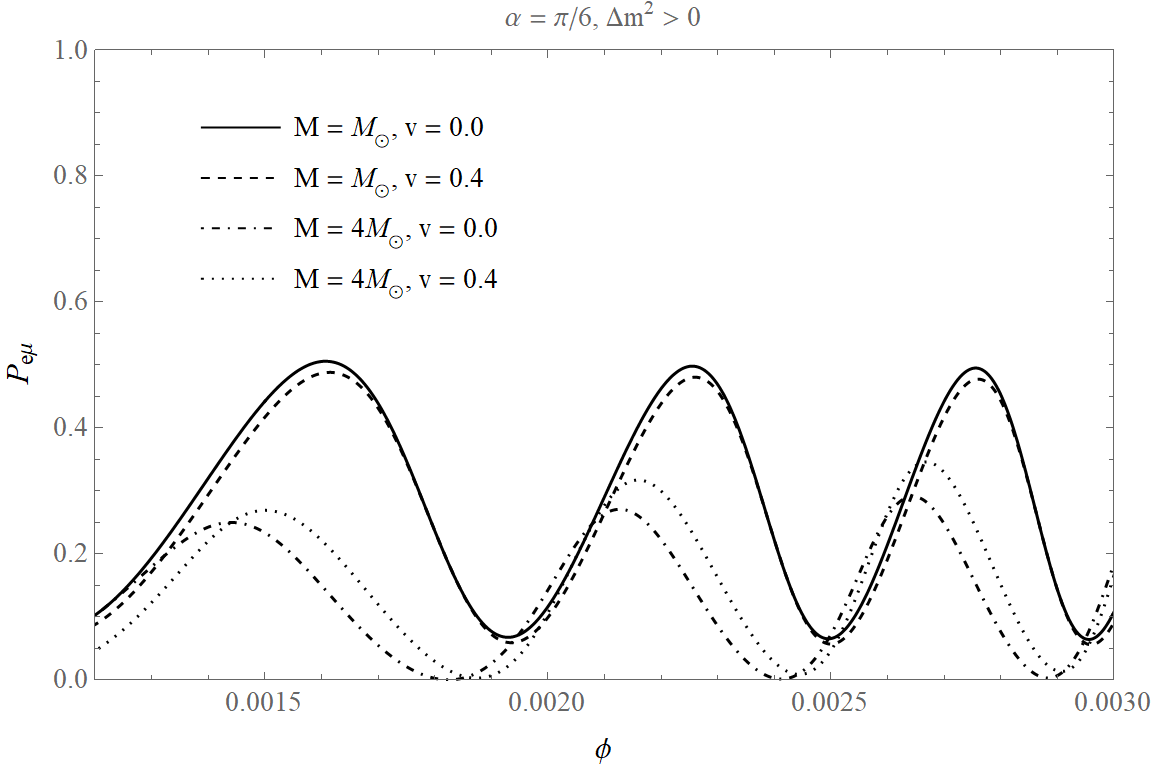}
        \includegraphics[width=0.45\textwidth]{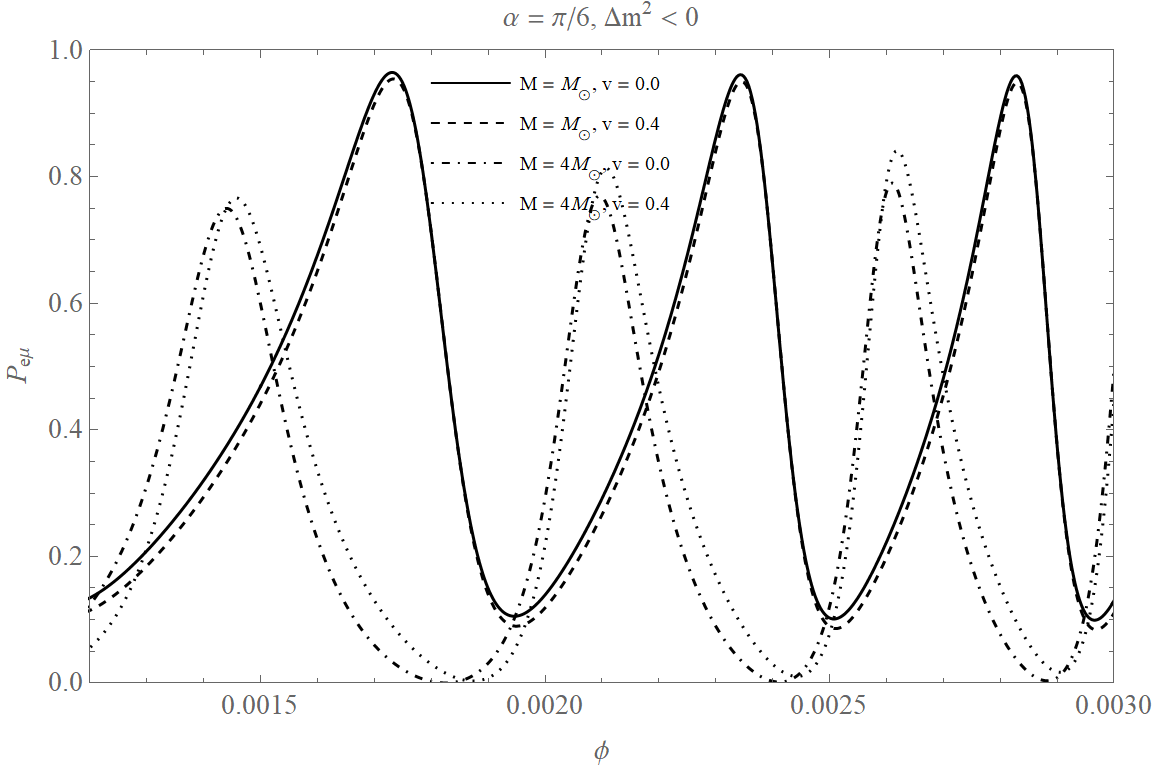}
    \end{center}
    \caption{ Left Panel: Neutrino oscillation probability for $ v = 0, M = M_\odot $ (solid curve), $ v = 0.4, M = M_\odot $ (dashed curve), $ v = 0, M = 4M_\odot $ (dotted-dashed curve) and $ v = 0.4, M = 4M_\odot $ (dotted curve) for normal hierarchy. Right panel: Neutrino oscillation probability for $ v = 0, M = M_\odot $ (solid curve), $ v = 0.4, M = M_\odot $ (dashed curve), $ v = 0, M = 4M_\odot $ (dotted-dashed curve) and $ v = 0.4, M = 4M_\odot $ (dotted curve) for inverted hierarchy. 
    Values of the other parameters are as follows: $ \left| \Delta m^2 \right| = 10^{-3} \rm{eV}^2 $, $ \alpha = \pi/6 $ and the lightest neutrino is massless. We use Eq.~\eqref{eq-pr} and \eqref{eq-norm} with \eqref{aijbpq} to plot these figures.  \label{fig5}}
\end{figure*}

In this section, we shall calculate the oscillation probability of neutrinos that are lensed by a KK black hole. We shall consider neutrinos with mass eigenstates $ \ket{\nu_i} $ travelling in the KK black hole geometry from a source $ S $ to a detector $ D $ through different classical paths. A flavour eigenstate $ \ket{\nu_\alpha} $ propagated from $ S $ to $ D $ through a path denoted by $ p $ is given by 
\begin{equation}
    \ket{\nu_\alpha(t_D,x_D)} = N \sum_i U_{\alpha i}^* \sum_p \exp \left( -i\Phi_i^p \ket{\nu_i (t_S,x_S)} \right),
\end{equation}
where $ \Phi_i^p $ is the phase of neutrino oscillation with the impact parameter $ b_p $ now being a path-dependent parameter and $ N $ is a normalization factor. We suppose that a neutrino is produced in a flavor eigenstate $ \alpha $ at the source $ S $ and detected in a flavor eigenstate $ \beta $ at the detector $ D $. Then the probability of oscillation is given by 
\begin{equation}\label{prob-1}
    \begin{aligned}
        \mathcal{P}_{\alpha\beta}^{lens} &= |\braket{\nu_\beta |\nu_\alpha(t_D,x_D)}|^2  \\
        &= |N|^2 \sum_{i,j}U_{\beta i}U_{\beta j}^*U_{\alpha j}U_{\alpha i}^* \sum_{p,q} \exp \left( 
        -i\Delta \Phi_{ij}^{pq} \right),
    \end{aligned}
\end{equation}
where $N$ is given by
\begin{equation}
    |N|^2 = \left( \sum_i |U_{\alpha i}|^2 \sum_{p,q} \left( -i\Delta \Phi_{ii}^{pq} \right) \right)^{-1}.
\end{equation}
$ \Delta \Phi_{ij}^{pq} $ is the phase difference between the paths $p$ and $q$ which is given by
\begin{equation}\label{phiAB}
    \Delta\Phi_{ij}^{pq} = \Phi_p^i - \Phi_q^j = \Delta m_{ij}^2 A_{pq} + \Delta b_{pq}^2 B_{ij},
\end{equation}
with
\begin{equation}\label{aijbpq}
    \begin{aligned}
        &A_{pq} = \frac{r_S + r_D}{2E_0}\left[ 1 + \left( 1 + \frac{Q^2}{2M^2} \right)\frac{2M}{r_S+r_D} - \frac{\sum b_{pq}^2}{4r_Sr_D} \right], \\
        &B_{ij} = - \frac{\sum m_{ij}^2}{8E_0}\left( \frac{1}{r_S} + \frac{1}{r_D} \right).
    \end{aligned}
\end{equation}
In the above set of equations, the quantities $ \Delta b_{pq}^2 $, $ \sum b_{pq}^2 $, $ \Delta m_{ij}^2 $ and $ \sum m_{ij}^2 $ are given by
\begin{equation}
    \begin{aligned}
        &\Delta b_{pq}^2 = b_p^2 - b_q^2, \\
        &\sum b_{pq}^2 = b_p^2 + b_q^2, \\
        &\Delta m_{ij}^2 = m_i^2 - m_j^2, \\
        &\sum m_{ij}^2 = m_i^2 + m_j^2.
    \end{aligned}
\end{equation}
We can clearly notice from the above expressions that $ A_{pq} $ and $ B_{ij} $ are invariant under the change of their respective indices, i.e. $ A_{pq} = A_{qp} $ and $ B_{ij} = B_{ji} $. This suggests that under a simultaneous change of both of its indices, the phase difference changes sign $ \Delta\Phi_{pq}^{ij} = -\Delta\Phi_{qp}^{ji} $. For those neutrino paths for which $ p = q $, we have that $ \Delta b_{pq}^2 $ vanishes and the oscillation probability only depends on the difference of squared masses of neutrinos. However, for general paths, $ p \neq q $, the oscillation probability also depends on the sum of squared masses. Therefore, in principle, gravitationally lensed neutrinos can provide information about the individual neutrino masses if the spacetime properties are known from other observations \cite{Swami:2020qdi}.

Now we would like to consider a simple toy model of two neutrino flavors to understand how the boost velocity $ v $ affects the oscillation probability. Specifically, we shall evaluate
the two-flavor neutrino oscillation probability at a generic point in a plane connecting the source $ S $, the lens and the detector $ D $ in the weak field limit. We substitute Eq.\eqref{phiAB} in Eq.\eqref{prob-1} to get
\begin{equation}
    \begin{aligned}
        \mathcal{P}_{\alpha\beta}^{lens} &= |N|^2\Bigg[ \sum_{i,j}U_{\beta i}U_{\beta j}^*U_{\alpha j}U_{\alpha i}^*\Big( \sum_{p=q} \exp \left( -i\Delta m_{ij}^2 A_{pp} \right)  \\
        & \quad + 2\sum_{p>q} \cos \left( \Delta b_{pq}^2 B_{ij} \right) \exp \left( -i\Delta m_{ij}^2A_{pq} \right) \Big) \Bigg],
    \end{aligned}
\end{equation}
where
\begin{equation}
    |N|^2 = \left( N_{path} + \sum_{i}|U_{\alpha i}|^2 \sum_{q>p} 2 \cos \left( \Delta b_{pq}^2 B_{ii} \right) \right)^{-1}.
\end{equation}
For simplicity, we consider the neutrinos to be propagating in the equatorial plane ($ \theta = \pi/2 $) and, in this case, $ N_{path} = 2 $. To understand the probabilities qualitatively and quantitatively, we consider a transition from electron to muon neutrino, $ \nu_e \rightarrow \nu_\mu $. For this process, the probability can be written as
\begin{equation}\label{eq-pr}
    \begin{aligned}
        \mathcal{P}_{e\mu}^{lens} &= |N|^2 \sin^2 2\alpha \Bigg[ \sin^2\left( \Delta m^2 \frac{A_{11}}{2}\right) + \sin^2\left( \Delta m^2 \frac{A_{22}}{2}\right) \\
        & \quad - \cos \left( \Delta b^2 B_{12} \right)\cos \left( \Delta m^2 A_{12} \right) + \frac{1}{2} \cos \left( \Delta b^2 B_{11} \right)  \\
        & \quad + \frac{1}{2}\cos \left( \Delta b^2 B_{22} \right)   \Bigg],
    \end{aligned}
\end{equation}
where
\begin{equation}\label{eq-norm}
    |N|^2 = \left[ 2\left( 1 + \cos^2\alpha \cos(\Delta b^2 B_{11}) + \sin^2 \alpha \cos (\Delta b^2 B_{22}) \right) \right]^{-1}.
\end{equation}
In these expressions, $ \Delta b^2 = \Delta b_{12}^2 $, $ \Delta m^2 = \Delta m_{21}^2 $, $ \alpha $ is the mixing angle and $ A_{pq} $ and $ B_{ij} $ are given by Eq.\eqref{aijbpq}. We would like to express the path-dependent impact parameter $b_p$ in terms of the geometric quantities of the system. Figure \ref{fig1} shows the schematic representation of the geometric system of lensing phenomena. Neutrinos are produced at the source $ S $, get lensed by a KK black hole metric and are later detected at $ D $. The physical distances from the source to the lens and from the lens to the detector are $ r_S $ and $ r_D $, respectively in the $ (x,y) $ coordinate system. Let us consider another coordinate system $ (x',y') $ obtained by rotating the $ (x,y) $ system by an angle $ \varphi $. These two systems then would be related by the relations $ x' = x \cos \varphi + y \sin \varphi  $ and $ y' = -x \sin \varphi + y \cos \varphi $. Now, the angle of deflection $\delta$ of neutrinos in the rotated frame can be given by
\begin{equation}\label{delta}
    \begin{aligned}
        \delta = \frac{y_D' - b}{x_D'} &= -\frac{4m}{b}\left( 1 + \frac{v^2}{2(1-v^2)} \right) = -\frac{4M}{b} = -\frac{2R_x}{b},
    \end{aligned}
\end{equation}
where $ R_x $ is just $ 2 M $ and $ M $ is the physical mass of the source\footnote{Notice that equation \eqref{delta} provides the deflection angle for weak lensing of null rays, which includes photons. Therefore just by measuring the deflection angle in the weak field one would not be able to discriminate between different metrics.}. Here $ (x_D', y_D') $ is the location of the detector. Now from Figure \ref{fig1}, we use the identity $ \sin \varphi = b/r_S $ and the above equation becomes
\begin{equation}\label{polynom}
    \left( 2R_x x_D + b y_D \right)\sqrt{1-\frac{b^2}{r_S^2}} = b^2\left( \frac{x_D}{r_S} + 1  \right) - \frac{2R_x b y_D}{r_S}.
\end{equation}
Assuming a nearly-circular trajectory of the detector around the lens, we have $ x_D = r_D \cos\phi $ and $ y_D = r_D \sin \phi $. Now, Eq.\eqref{polynom} is a quartic polynomial and it can be solved to obtain the impact parameters in terms of $ r_S $, $ R_x $ and the lensing location ($ x_D, y_D $). For a quantitative understanding, we consider a Sun-Earth system with the Sun as the lens and the Earth as a detector. In this example, we shall use the typical values of the geometrical quantities in the solar-system, assuming that the gravitational field of the Sun is represented by the exterior of the KK metric in the weak field limit. We assume the source is behind the Sun and it emits high-energy neutrinos with $ E_0 = 10 \rm{MeV} $. We consider $ r_D = 10^8 \rm{km} $ and $ r_S = 10^5r_D $ and numerically solve Eq.\eqref{polynom} and obtain two real roots $b_1$ and $b_2$ for each $ \phi $. Physical mass of the lens is considered to be $ M = 1-5 M_\odot $ and neutrino mass squared difference is $ |\Delta m^2| = 10^{-3} \rm{eV}^2 $. With these numerical values we plot the oscillation probability of neutrinos using Eq.~\eqref{eq-pr}, \eqref{eq-norm} and \eqref{aijbpq} with respect to the azimuth angle $ \phi $.

\begin{figure*}
    \begin{center}
        \includegraphics[width=0.45\textwidth]{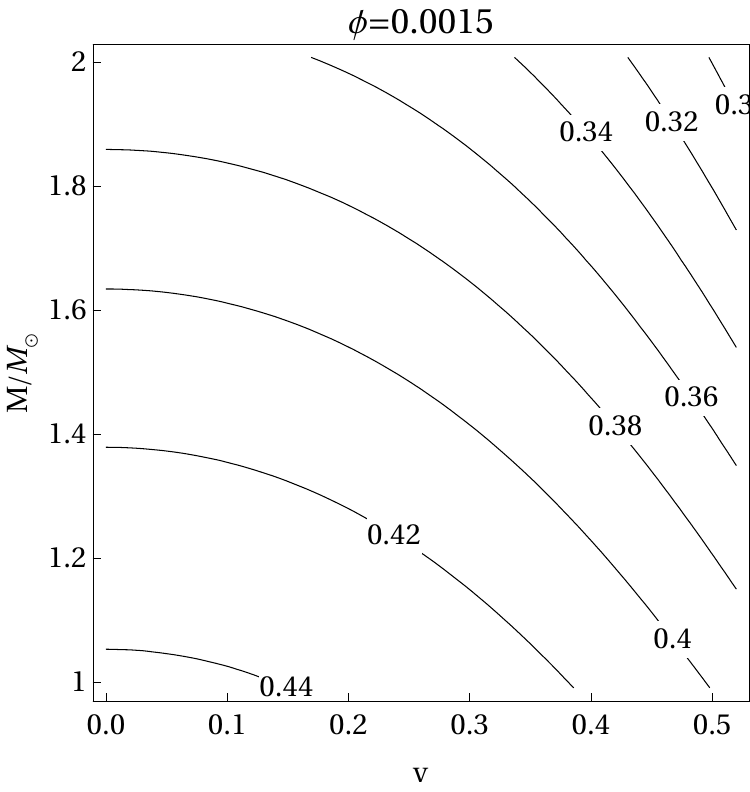}
        \hspace{0.5cm}
        \includegraphics[width=0.45\textwidth]{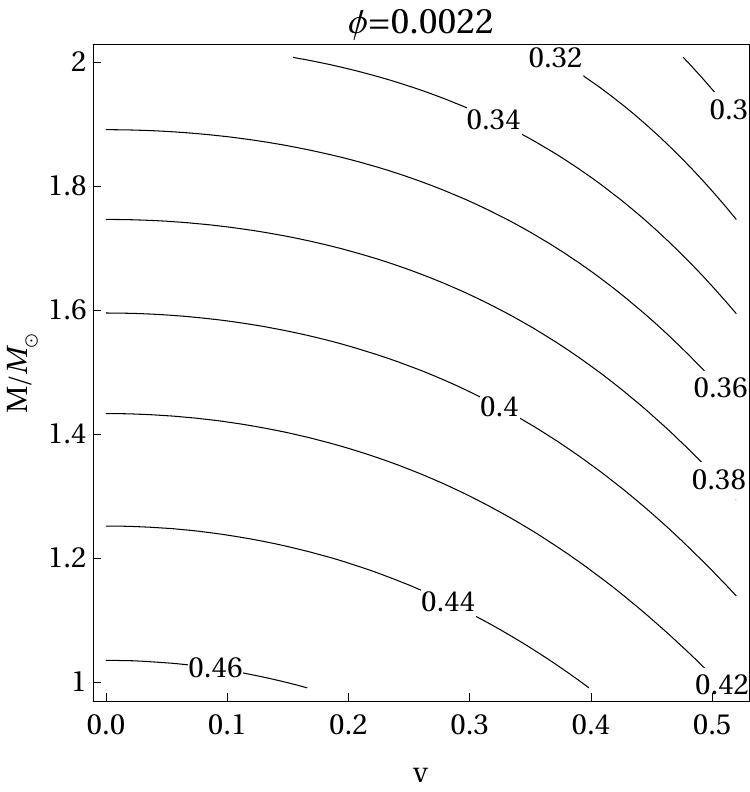}
    \end{center}
    \caption{The degeneracy between the determination of the mass parameter $M$ and the boost parameter $v$ for a given probability $\mathcal{P}_{e\mu}$ at a fixed angle $\phi$ is illustrated by the 2D contour plot of the implicit function $M(v)$ obtained from $\mathcal{P}_{e\mu} (M,v)=\rm{const}.$, namely the plot of the values of the boost parameter $ v $ and the lens mass $ M $ which give the same probability. Each curve corresponds to a fixed value of $\mathcal{P}_{e\mu}$ as shown on the line itself. The left panel shows the contour plots for $\phi=0.0015$, while the right panel shows the contour plots $M(v)$ for $\phi=0.0022$. Keep in mind that the probability also changes with the angle, therefore for the same source the curves $M(v)$ would correspond to different $\mathcal{P}_{e\mu}$ at different angles. \label{fig6}}
\end{figure*}


The oscillation probability for the two flavor toy model of neutrinos is shown in Figures \ref{fig2}, \ref{fig3}, \ref{fig4} and \ref{fig5} as a function of the the azimuth angle $\phi$ that signifies the detector's position in the orbit. Figure \ref{fig2} shows the oscillation probability of gravitationally lensed neutrinos in a Schwarzschild spacetime, ($ v = 0 $). The solid and the dashed line corresponds to normal hierarchy ($ \Delta m^2 > 0 $) and inverted hierarchy ($ \Delta m^2 < 0 $) of neutrinos respectively. Verifying the results of \cite{Swami:2020qdi,Chakrabarty:2021bpr}, we can see that the probability in the case of inverted hierarchy is quite large for particular values of $\phi$. 

We see similar behavior also in case of non-zero $v$, i.e. in a KK black hole spacetime, in Figure \ref{fig3} and \ref{fig4}. 
From Figure \ref{fig4} we can see that for sufficiently large values of the boost parameter $ v $, the probability deviates significantly from that of the Schwarzschild case, suggesting that the detection of a sufficiently large number of neutrinos could provide a way to test the geometry. The solid line there shows the probability of oscillation for $v=0$ (Schwarzschild case), and the dashed and the dotted-dashed line shows the oscillation probability for $ v = 0.2 $ and $ v = 0.4 $ respectively. This is true for both normal and inverted hierarchy.  

Next, we investigate the effect of the mass of the lens and the boost parameter together in Figure \ref{fig5}. The left and right panels of Figure \ref{fig5} show the probability of oscillation of neutrinos for normal ($\Delta m^2 > 0$) and inverted hierarchy ($\Delta m^2 < 0$) for two different masses of the lens. The solid and dashed line corresponds to a solar mass lens with $ v = 0 $ and $ v = 0.4 $, respectively. The dotted and dotted-dashed line corresponds to a $ 4M_\odot $ lens with $ 
v = 0 $ and $ v = 0.4 $, respectively. We can see that a more massive lens can create a bigger shift in the oscillation probability. In certain cases, the shift in probability for a $ 4M_\odot $ gravitational lens is almost double that of a $ M_\odot $ lens. We can see that the effect of the lens mass on the probability of flavor transition is more significant than the one of the boost parameter. This can be understood by looking at the second term under the brackets of the first equation of Eq.~\eqref{aijbpq} which is responsible for the effects of the mass and the boost parameter. Using the relation $ Q/M = 2v/(2-v^2) $ we can express the term as 
\begin{equation}
    \begin{aligned}
        \left( 1 + \frac{Q^2}{2M^2} \right)\frac{2M}{r_S+r_D} &= \left( 1 + \frac{2v^2}{(2-v^2)^2} \right)\frac{2M}{r_S+r_D} \\
        &\simeq \left( 1 + \frac{v^2}{2} \right)\frac{2M}{r_S+r_D}.
    \end{aligned}    
\end{equation}
The numerical values of the physical mass $ M $ used in the plots are one order of magnitude higher than that of the boots parameter $ v $ which has strict limits by definition ($0<v<1$). This difference is then carried to the probability expression Eq.~\eqref{eq-pr} where $ M $ and $ v $ appears under the trigonometric functions. Therefore the observed effect for lens mass is much higher than that of the boost parameter. 

Finally we investigated the degeneracy for the measurements of the lens mass and boost parameter simultaneously. In Figure \ref{fig6} we plotted the implicit function $M(v)$ obtained at a given angle $\phi$ for a fixed value of the probability $\mathcal{P}_{e\mu}$. We see that for any given value of $\mathcal{P}_{e\mu}$ there are several values of $M$ and $v$ that produce the same probability. 
This shows that, similarly to what happens with other kinds of observations, one needs multiple independent measurements of one of the quantities in order to break the degeneracy and determine $M$ and $v$. In fact, obtaining the degeneracy plot $M(v)$ for a different kind of measurement, such as for example accretion disk's spectroscopy, would allow to determine the values of $M$ and $v$. This kind of degeneracy happens for most measurements of black hole quantities when one wishes to compare with alternative models characterized by additional parameters and is the reason why independent measurements of the same quantity are important.

Nevertheless, it is worth mentioning that the implicit plot changes with $\phi$, which is the detector's azimuthal angle, which in turn, for Earth, depends on the time of the year. Therefore by tracking the implicit plots for $M(v)$ at different angles which correspond to the probability $\mathcal{P}_{e\mu}$ at that angle one may be able to constrain the range of allowed values of $M$ and $v$ for the source. In principle, by overlapping the obtained range with another range of possible values of $M$ for the same source, as obtained through a different method (for example, x-ray spectroscopy), one could be able to further constrain $M$ and therefore $v$, thus testing the nature of the geometry.

\section{Discussion}\label{7}

We considered gravitational lensing of neutrinos by a non rotating KK black hole as a possible tool to estimate the properties of the geometry and potentially distinguish such hypothetical objects from black holes in classical GR. We showed how the boost parameter $v$ of the KK black hole affects the probability of neutrino oscillations and showed that there exist a degeneracy between in the measurement of the boost parameter and gravitational mass of the object using this method. As a consequence, the observation of a black hole candidate with this method would provide only a range for the allowed values of the object's mass and boost parameter and would not be able to rule out the validity of the KK metric for the source.
It is well known that similar degeneracies arise for the measurement of the properties of black hole candidates also from other methods such as gravitational waves, spectroscopy or shadow (see for example \cite{Joshi:2013dva,Bambi:2013hza,Abdikamalov:2019ztb} for specific examples or \cite{Berti:2015itd, Cardoso:2016ryw} for a more general discussion of tests of GR). We argued that an independent measurement of the object's mass with two separate methods would allow to reduce the allowed range of values for $v$ and thus constrain the validity of the KK metric as the exterior field of a compact object.
We emphasize again that the gravitational lensing of neutrinos can provide information about the boost parameter even in the weak-field limit. On the other hand, the lensing of photons in the weak-field limit only depends on the physical mass of the black hole and thus is unable to discriminate between different geometries with the same $M$.

In our analysis we focused on the non rotating case, however for non vanishing spin parameter we would expect that the effects of (slow) rotation, at least to lowest order, would be similar to what happens in the Kerr case, as discussed for example in \cite{Swami:2022xet}.

It is worth mentioning that the above tools need not be restricted to testing black hole candidates. For example, regarding neutron stars, neutrino lensing observations along with other astrophysical probes can be used to determine physical properties of the system such as the mass and the size. Therefore, in principle it may be possible to distinguish between a black hole and a neutron star. 
On the other hand other types of tests, such as for example, the probe of dark matter halos, may be impossible due to the fact that for neutrinos propagating inside a medium the oscillations would depend on both gravity and matter effects and the two effects would not easily be disentangled.

Note that, we performed our study of neutrino oscillations in the plane wave approximation. In a realistic scenario, a wave packet approach is more appropriate as the neutrinos are produced and detected as wave-packets of finite width in position space. The wave-packet approach introduces a new length scale, beyond which the transition probability between different flavours saturates to a value depending only on the leptonic mixing parameters. This phenomena is known as decoherence \cite{Giunti:1991ca,Giunti:1997sk}. 
It has been shown that the decoherence length depends explicitly on the mass parameter of the Schwarzschild spacetime in gravitational lensing of neutrinos and hence monitoring of the decoherence can provides an avenue for estimates of the spacetime parameters \cite{Swami:2021wbf,Chatelain:2019nkf}. 
However, the coherence condition suggests that to maintain coherence over distances of the order of the Sun-Earth system, one would require very high precision in determining the energies or momenta of particles involved in the production and detection processes of neutrinos \cite{Swami:2020qdi,Swami:2021wbf}. A more detailed discussion on decoherence in modified metrics is left for future projects.

In general, neutrino mass matrix and mixing angles can be modified by Yukawa-type coupling between fermions and the dilaton field $\varphi$ to include additional components that change in time periodically with a frequency and amplitude determined by the mass and energy density of the dilation field, see for an example in \cite{Berlin:2016woy}. In connection with this type of models, solar neutrino detectors such as Super-K and SNO provide a particularly interesting class of investigations. However, null results from recent searches for anomalous periodicities in the solar neutrino flux lead to an upper bound on the Yukawa coupling of the dilaton field which is very small. Thus, we can ignore this effect in our analysis.
Additionally, it has been suggested \cite{Damour:1994zq} that Yukawa-type coupling of the dilaton field which gives an additional contribution to the usual graviton exchange gravity may lead to the violation of the equivalence principle (VEP). Testing of VEP effect in neutrino oscillation experiments has been discussed in \cite{Halprin:1997na,Klapdor-Kleingrothaus:2000yox}. Further studies on how to evaluate this effect in our situation could be interesting.

As of now almost all observations of any black hole candidate have been done using only one tool and the precision of the data is in most cases not sufficient to provide stringent constraints on the metric parameters. 
Similarly, the detection of neutrinos from galactic and extra-galactic astrophysical sources is still in its infancy \cite{IceCube:2018cha,IceCube:2018dnn} and therefore the methods discussed here are not experimentally applicable today. However, 
the recent results on galactic neutrinos from the IceCube Neutrino Observatory \cite{Abbasi:2023bvn} suggest that this may soon change and
we believe that in the future, when more data and increased precision will be available, neutrino oscillations may become an important tool for our understanding of the nature of extreme compact objects.

\acknowledgments
This work was supported in part by the NSRF via the Program Management Unit for Human Resources and Institutional Development, Research and Innovation [grant number B05F650021]. A.C. is also supported in part by Thailand Science research and Innovation Fund Chulalongkorn University (IND66230009). T.T. is supported by School of Science, Walailak University, Thailand. DM and HC acknowledge support from Nazarbayev University Faculty Development Competitive Research Grant No. 11022021FD2926. HC would also like to thank AC for the kind hospitality during the 10th Bangkok Workshop on High-energy Theory.



\bibliographystyle{apsrev}
\bibstyle{apsrev}
\bibliography{ref,ref1}
\end{document}